\documentclass[aps,nofootinbib]{revtex4}
\def\be {\begin{equation}}
\def\ee {\end{equation}}
\def\ba {\begin{eqnarray}}
\def\ea {\end{eqnarray}}
\def\nn {\nonumber}
\def\a  {\alpha}
\def\b  {\beta}

\def\D  {\Delta}

\def\l  {\lambda}

\def\m  {\mu}

\def\O  {\Omega}
\def\p  {\pi}

\def\r  {\rho}

\def\s {\sigma}
\def\t  {\tau}
\def\la {\label}
\def\le {\left}
\def\ri {\right}
\def\pa {\partial}
\def\f {\frac}

\def\no {\noindent}
\def\bi {\begin{itemize}}
\def\ei {\end{itemize}}

\def\vs {\vspace}

\def\ul {\underline}

\def\laq{\hbox{~}\raise 0.4ex\hbox{$<$}\kern -0.8em\lower 0.62ex\hbox{$\sim$}\hbox{~}}
\def\gaq{\hbox{~}\raise 0.4ex\hbox{$>$}\kern -0.7em\lower 0.62ex\hbox{$\sim$}\hbox{~}}
\begin{document}

\title{How classical are TeV-scale black holes?}
\author{Marco Cavagli\`a}

\affiliation{Dept. of Physics and Astronomy, The University of Mississippi,\\
PO Box 1848, University, Mississippi 38677-1848, USA}
\email{cavaglia@phy.olemiss.edu}

\author{Saurya Das}
\affiliation{Dept. of Physics, University of Lethbridge,\\
4401 University Drive, Lethbridge, Alberta T1K 3M4, CANADA}
\email{saurya.das@uleth.ca}

\begin{abstract}
We show that the Hawking temperature and the entropy of black holes are
subject to corrections from two sources: the generalized uncertainty
principle and thermal fluctuations. Both effects increase the temperature
and decrease the entropy, resulting in faster decay and ``less classical''
black holes. We discuss the implications of these results for TeV-scale black
holes that are expected to be produced at future colliders. 
\end{abstract}

\maketitle

\section{Introduction}

The possibility of the existence of large extra dimensions has recently opened
up new and exciting avenues of research in quantum gravity \cite{add,rs}. In
particular, a host of interesting work is being done on different aspects of
low-energy scale {\em quantum gravity phenomenology}. One of the most
significant sub-fields is the study of black hole (BH) \cite{bhs} and brane
\cite{branes} production at particle colliders, such as the (Very) Large Hadron
Collider [V(LHC)] \cite{lhc} and the muon collider \cite{mucol}, as well as in
ultrahigh energy cosmic ray (UHECR) airshowers \cite{uhecr}. (For recent
reviews, see Refs.~\cite{reviews}.) Several scenarios predict the fundamental
Planck scale to be of order of the TeV. The simplest model postulates a number
$n$ of toroidally compactified extra dimensions with length ranging from a few
microns ($n=2$) to a  Fermi ($n=7$) \cite{add}. Extra dimensions of infinite
size and non-trivial ``warp-factor'' may also lead to similar predictions
\cite{rs}. In either case, particle collisions with center-of-mass (c.m.)
energy above the fundamental Planck scale, and impact parameter smaller than
the horizon radius corresponding to that energy, should produce BHs and branes
\cite{formation}. (For criticisms, however, see Refs.~\cite{criticisms}.) Since
the c.m.~energy of next-generation particle colliders and UHECR primaries is as
high as tens or hundreds of TeV, BH and brane production is likely to be
observed. For this kind of event, the initial mass of the gravitational object
is expected to be of the order of a few Planck masses.

Newly formed BHs first lose hair associated with multipole and angular momenta,
then approach classically stable Schwarzschild solutions, and finally evaporate
via Hawking radiation \cite{evaporation}. Decay time and entropy completely
determine the observables of the process. BH formation and decay can be
described semiclassically, provided that the entropy is sufficiently large. The
timescale for the complete decay of a BH to up to its supposed final
Planck-sized remnant is expected to be of order of the TeV$^{-1}$. 

BH thermodynamic quantities depend on the Hawking temperature $T_H$ via the
usual thermodynamic relations (Stefan-Boltzmann law). The Hawking temperature
undergoes corrections from many sources, and these corrections are particularly
relevant for BHs with mass of the order of the Planck mass. Therefore, the
study of TeV-scale BHs in UHECR and particle colliders requires a careful
investigation of how temperature corrections affect BH thermodynamics. In this
article, we concentrate on the corrections due to the generalized uncertainty
principle (GUP) and thermal fluctuations of thermodynamic systems. These
corrections are not tied down to any specific model of quantum gravity; the GUP
can be derived using arguments from string theory \cite{gup-strings} as well as
other approaches to quantum gravity \cite{gup-others,Maggiore:rv}. Similarly,
corrections from thermal effects do not depend on the underlying quantum
gravity theory since thermal fluctuations are present in any system. This
generality provides in fact a strong motivation in studying GUP and thermal
fluctuation effects. 

We show below that the BH decay rate is increased by GUP and thermal
fluctuation corrections, resulting in shorter decay times. Thus the BHs may not
behave like well-defined resonances. We also show that a diminished entropy
leads to smaller particle emission during the evaporation phase and ``less
classical'' BHs. The paper is organised as follows. In the next section, we
review the connection between the uncertainty principle and Hawking radiation
\cite{Adler:2001vs,Cavaglia:2003qk}. In section \ref{sectiongup}, we derive
corrections to the Hawking decay rate, entropy, and multiplicity due to the
GUP. We show that for BHs with mass $M=5$ - $10 M_{Pl}$, the decay time and the
entropy (or multiplicity) dramatically decrease if the GUP parameter is
nonvanishing. In section \ref{entcorr} we review the corrections to
thermodynamic quantities due to thermal fluctuations. We apply these results
to  BHs in section \ref{fluct}, where we show that BH decay time and entropy
also decrease compared to their semiclassical value. We conclude with a summary
of our results and a brief discussion of open questions in section \ref{concl}.
\section{Uncertainty Principle and Hawking radiation}
In this section we review and generalize to $d$ dimensions the derivation of
the Hawking radiation of Adler {\it et al.} \cite{Adler:2001vs}. A
$d$-dimensional spherically symmetric BH of mass $M$ (to which the collider BHs
will settle into before radiating) is described by the metric
\be
ds^2 = -
\left(1 - \f{16\p  G_d M}{(d-2)\O_{d-2}c^2 r^{d-3}} \right) c^2 dt^2
+ \left(1 - \f{16\p  G_d M}{(d-2)\O_{d-2}c^2 r^{d-3}} \right)^{-1} dr^2
+ r^2 d\O_{d-2}^2\,,
\la{sc1}
\ee
where $\O_{d-2}$ is the metric of the unit $S^{d-2}$ and $G_d$ is the
$d$-dimensional Newton's constant. Since the Hawking radiation is a quantum
process, the emitted quanta must satisfy the Heisenberg uncertainty principle
\be
\D x_i\D p_j\gaq \hbar\delta_{ij}\,,
\la{up}
\ee
where $x_i$ and $p_j$, $i,j=1\dots d-1$, are the spatial coordinates and
momenta, respectively. By modelling a BH as a $(d-1)$-dimensional cube
of size equal to twice its Schwarzschild radius $r_s$, the uncertainty in the
position of a Hawking particle at the emission is
\be
\D x \approx 2r_s = 2\l_d \le[\f{G_d M}{c^2}\ri]^{1/(d-3)}\,,
\la{Dx}
\ee
where $\l_d=\le[16\p/((d-2)\O_{d-2})\ri]^{1/(d-3)}$. Using Eq.\ (\ref{up}), the
uncertainty in the energy of the emitted particle is
\be
\D E \approx c \D p\approx \f{M_{Pl}c^2}{2\l_d} m^{-1/(d-3)}\,,
\la{DE}
\ee
where $m=M/M_{Pl}$ is the mass in Planck units and
$M_{Pl}=[\hbar^{d-3}/c^{d-5}G_d]^{1/(d-2)}$ is the $d$-dimensional Planck mass.
$\D E$ can be identified with the characteristic temperature of the BH
emission, i.e.\ the Hawking temperature. Setting the constant of proportionality to
$(d-3)/2\p$ we get
\be
T_H = \f{d-3}{4\p\l_d}~M_{Pl}c^2~m^{-1/(d-3)}\,.
\la{haw}
\ee
The energy radiated per unit time is governed by the Stefan-Boltzmann law. The
surface gravity is constant over the horizon. Thus the Hawking temperature of
the higher-dimensional BH and the temperature of the induced BH on the brane
are identical. The BH temperature $T_H$ can be used in the calculation of the
emission rate. Neglecting thermal emission in the bulk, and assuming that the
brane has $\cal D$ spacetime dimensions (we will substitute ${\cal D}=4$ at the
end, since BHs are supposed to radiate mainly on the brane
\cite{evaporation}),  the emission rate for a massless scalar particle on the
brane is
\be
\f{dM}{dt} = - {c \s_{{\cal D}}} A_{{\cal D}} T^{{\cal D}}\,,
\label{dMdt}
\ee
where $A_{{\cal D}}=\O_{{\cal D}-2} r_c^{{\cal D}-2}$ is the horizon area of
the induced BH with radius $r_c=[(d-1)/2]^{1/(d-3)}[(d-1)/(d-3)]^{1/2}r_s$, and
\be
\s_{{\cal D}} = \f{\O_{{\cal D}-3} \Gamma({\cal D}) \zeta({\cal D})}
{({\cal D}-2)(2\p \hbar c)^{{\cal D}-1}}
\equiv \f{{\bar \s}_{\cal D}}{(\hbar c)^{{\cal D}-1}}
\ee
is the Stefan-Boltzmann constant in ${\cal D}$-spacetime dimensions. If the BH
evaporates into different particle species on the brane, the Stefan-Boltzmann
constant has to be multiplied by the factor
\be
\sum_i c_i({\cal D})\Gamma_{s_i}({\cal D})f_i({\cal D})\,,
\ee
where the sum is over all particle flavors, $c_i$ are the degrees of freedom of
the species $i$, $\Gamma_{s_i}$ are the greybody factors for spin $s_i$ and
$f_i=1$ ($f_i=1-2^{1-{\cal D}}$) for bosons (fermions). (We neglect the energy
dependence of the greybody factors. See, e.g., Refs.~\cite{evaporation}.)
Expressing Eq.~(\ref{dMdt}) in terms of $m$, we obtain
\be
\f{dm}{dt}=-\f{\m}{t_{Pl}}~m^{-2/(d-3)}\,,
\la{rate1}
\ee
where $t_{Pl}=(\hbar G_d/c^{d+1})^{1/(d-2)}$ is the Planck time, and
\be
\mu = \left(\frac{r_c}{r_s}\right)^{{\cal D}-2}
\left(\frac{d-3}{4\p}\right)^{{\cal D}}\frac{{\bar \s}_{{\cal D}}\O_{{\cal
D}-2}}{\l_d^2}\,.
\la{mu}
\ee
Integration over $t$ yields the decay time
\be
\t_0 =  \m^{-1} \le(\f{d-3}{d-1} \ri)m_i^{(d-1)/(d-3)}~t_{Pl}\,,
\la{decayt}
\ee
where $m_i\equiv M_i/M_{Pl}$, and $M_i$ is the initial BH mass. The decay time
$\tau_0$ is finite. Equation (\ref{rate1}) implies that the end stage of
Hawking evaporation is catastrophic, with infinite radiation rate and infinite
temperature. However, a heuristic argument suggests that the final temperature
and radiation rate are finite. At the last stage of evaporation, $\D E$ in
Eq.~(\ref{DE}) must be of the order of the BH mass, and $\D E=\D M
c^2\approx M_{end}c^2$. This implies a minimum BH mass $M_{end}\approx
M_{Pl}$, a maximum Hawking temperature $T_{max}={\cal O}(M_{Pl})$, and a
smaller decay time.

The thermodynamic properties of the BH can be computed via the usual
thermodynamic relations. The entropy and the BH specific heat are
\be
S_0=\frac{4\pi\l_d}{d-2}m^{(d-2)/(d-3)}=\frac{d-3}{d-2}\frac{Mc^2}{T_H}\,,
\la{s0}
\ee
and
\be
{\cal C}_0=-4\p\l_d m^{(d-2)/(d-3)}\,,
\la{c0}
\ee
respectively. The statistical total number of quanta emitted during the
evaporation is proportional to the initial entropy of the BH. The exact
relation is 
\begin{equation}
N=
S_0\frac{\zeta({\cal D}-1)}{({\cal D}-1)\zeta({\cal D})}
\frac{\sum_i c_i({\cal D})\Gamma_{s_i}({\cal D})f_i({\cal D}-1)}
{\sum_j c_j({\cal D})\Gamma_{s_j}({\cal D})f_j({\cal D})}\,.
\label{multi}
\end{equation}
The flavor multiplicity is
\begin{equation}
N_i=N\frac{c_i({\cal D})\Gamma_{s_i}({\cal D})f_i({\cal D}-1)}
{\sum_j c_j({\cal D})\Gamma_{s_j}({\cal D})f_j({\cal D}-1)}\,.
\label{multifla}
\end{equation}
Equation (\ref{multifla}) gives the statistical number of particles per
species produced during the evaporation process.
\section{Corrections to BH thermodynamics from the Generalized
Uncertainty Principle}
\la{sectiongup}
We now determine the corrections to the above results due to the GUP. The
general form of the GUP is
\be
\D x_i \geq \f{\hbar}{\D p_i}+\a^2 \ell_{Pl}^2 \f{\D p_i}{\hbar}\,,
\la{gup1}
\ee
where $\ell_{Pl}=(\hbar G_d/c^3)^{1/(d-2)}$ is the Planck length and $\a$ is a
dimensionless constant of order one. There are many derivations of the
GUP, some heuristic and some more rigorous. Equation (\ref{gup1}) can be
derived in the context of string theory \cite{gup-strings}, non-commutative
quantum mechanics \cite{gup-others}, and from minimum length
\cite{Garay:1994en} considerations \cite{Maggiore:rv}. The exact value of $\a$
depends on the specific model. The second term in r.h.s.\ of Eq.
(\ref{gup1}) becomes effective when momentum and length scales are of the
order of Planck mass and of the Planck length, respectively. This limit is
usually called ``quantum regime''. Inverting Eq.\ (\ref{gup1}), we obtain
\be
\f{\D x_i}{2\a^2\ell_{Pl}^2}\le[1-\sqrt{1-\f{4\a^2\ell_{Pl}^2}{\D x_i^2}}\ri]
\leq\f{\D p_i}{\hbar}
\leq\f{\D x_i}{2\a^2\ell_{Pl}^2}\le[1+\sqrt{1-\f{4\a^2\ell_{Pl}^2}{\D
x_i^2}}\ri]\,.
\la{gup2}
\ee
The left-inequality gives the correct $\ell_{Pl}/\D x_i \rightarrow 0$ limit
and will be considered henceforth. The GUP implies the existence of a minimum
length $L_{min}\approx\D x=2\a \ell_{Pl}$. The string regime and the classical
regimes are recovered by setting $\D x_i\approx 2\a \ell_{Pl}$ and $\D x_i\gg
\ell_{Pl}$ in Eq.\ (\ref{gup1}), respectively.

BHs with horizon radius smaller than $L_{min}$ do not exist. Therefore, the
minimum length implies the existence of a minimum BH mass
\be
M_{min}= \frac{d-2}{8\Gamma\left(\frac{d-1}{2}\right)}
\left(\alpha\sqrt{\pi}\right)^{d-3} \, M_{Pl}\,.
\label{minmass}
\ee
The minimum BH mass is a rapidly increasing function of the unknown parameter
$\alpha$ for $d\ge 6$; a value of $\alpha$ larger than unity may lead to a
minimum BH mass $M_{min}\gg M_{Pl}$.

The corrections to the BH thermodynamic quantities can be calculated by
repeating the argument of the previous section. Setting $\D x=2r_s$ the
GUP-corrected Hawking temperature is
\be
T_H' = \f{(d-3)\l_d}{2\p\a^2}~m^{1/(d-3)}
\le[ 1 - \sqrt{ 1 - \f{\a^2}{\l_d^2 m^{2/(d-3)}} }\ri]~M_{Pl}c^2\,.
\la{hawcorr1}
\ee
Equation (\ref{hawcorr1}) may be Taylor expanded around $\alpha=0$:
\be
T_H' = \f{(d-3)}{4\p \l_d}~m^{-1/(d-3)}~\le[1 +
\f{\a^2}{4\l_d^2 m^{2/(d-3)}}  + \cdots \ri]~M_{Pl}c^2\,.
\la{taylor1}
\ee
The GUP-corrected Hawking temperature is higher than the semiclassical Hawking
temperature $T_H$ of Eq.\ (\ref{haw}). The first-order correction is 
\be
\D T_{GUP} \equiv T_H' - T_H = \f{d-3}{16\p}~
\f{\a^2}{\l_d^3~m^{3/(d-3)}}~M_{Pl}c^2\,.
\la{delta1}
\ee
From the first law of BH thermodynamics the first-order correction to the BH
entropy is
\ba
\D S_{GUP} &=& - \f{\p \a^2 m^{(d-4)/(d-3)}}{(d-4)\l_d} ~~~d > 4\,,\nonumber \\
           &=& - \f{\p\a^2}{2}\ln(m) ~~~~~~~~~~~\,d=4\,.
\la{SGUP}
\ea
This follows from the exact expression:
\be
S_{GUP} = 2\p\l_d \le( \f{\a}{\l_d} \ri)^{d-2}~I(1,d-4,\l_d m^{1/(d-3)}/\a)\,,  
\la{exent}
\ee
where
\be
I(p,q,x) = \int_1^x dz z^q \le( z + \sqrt{z^2-1} \ri)^p\,.
\ee
From Eq.~(\ref{SGUP}) and Eq.~(\ref{exent}) it follows that the GUP-corrected
entropy is smaller than the semiclassical Bekenstein-Hawking.  The
GUP-corrected Stefan-Boltzmann law is
\be
\f{dm}{dt} =
- 2^{{\cal D}}\f{\m}{t_{Pl}}~m^{-2/(d-3)}~
\le[ 1 + \sqrt{1 - \f{\a^2}{\l_d^2~m^{2/(d-3)}} }\ri]^{-{\cal D}}\,.
\la{ratecorr1}
\ee
Taylor expanding Eq.~(\ref{ratecorr1}) we have
\be
\f{dm}{dt} =
- \f{\m}{t_{Pl} m^{2/(d-3)}}
\le[ 1 + \f{\a^2{\cal D}}{4\l_d^2 m^{2/(d-3)}} + \cdots \ri]\,.
\la{rate5}
\ee
The relative GUP first-order correction to the Stefan-Boltzmann law is positive:
\be
\D\le(\f{dm}{dt}\ri)/\le(\f{dm}{dt}\ri)_0=\f{\a^2{\cal D}}{4\l_d^2}
m^{-2/(d-3)}\,,
\ee
where $\le(dm/dt\ri)_0$ is defined in Eq.~(\ref{rate1}). The Hawking
evaporation ends at $m_{min}= M_{min}/M_{Pl}=  (\a/\l_d)^{d-3}$, where the
emission rate becomes imaginary. The emission rate is finite at the end:
\be
\le(\f{dm}{dt}\ri)_{m_{min}} = -2^{{\cal D}}\f{\m}{t_{Pl}}\le(\f{\l_d}{\a}\ri)^2\,.
\ee
This means that the end-point of Hawking radiation is not catastrophic. Since
the final emission rate is finite, it might be argued that once the final stage
has been reached, the BH evaporates completely by emitting a hard Planck-mass
quantum in a finite time ${\cal O}(t_{Pl})$. However, the BH specific heat
\begin{eqnarray}
{\cal C}\equiv T\f{\partial S}{\partial T}=-2\pi\l_d m^{(d-2)/(d-3)}\sqrt{1-
\frac{\alpha^2}{\l_d^2 m^{2/(d-3)}}}\cdot
\left(1+ \sqrt{1-
\frac{\alpha^2}{\l_d^2 m^{2/(d-3)}}}\right)
\la{sh}
\end{eqnarray}
vanishes at the endpoint. Therefore, the BH cannot exchange heat with the
surrounding space. The endpoint of Hawking evaporation in the GUP scenario is
characterized by a Planck-size remnant with maximum temperature
\begin{equation}
T_{\text{max}}=2T_0\Big|_{M=M_{\text{min}}}\,.
\end{equation}
The GUP prevents BHs from evaporating completely, just like the standard
uncertainty principle prevents the hydrogen atom from collapsing. The existence
of BH remnants as a consequence of the GUP was pointed in
Refs.~\cite{Adler:2001vs} in the context of primordial BHs in cosmology
\cite{primordial}. BH remnants have also been predicted in string and quantum
gravity models \cite{remnants} and could play an important role in cosmology.
(See, e.g., Refs.~\cite{cosmicbhs}.)

The GUP implies a faster BH decay. The first-order decay time is
\be
\t_1 = \m^{-1} \le( \f{d-3}{d-1} \ri)
\le\{ \le[m_i^{(d-1)/(d-3)} -  \f{{\cal D} (d-1)\a^2}{4(d-3) \l_d^2}~m_i \ri]
- \le[1  -  \f{{\cal D} (d-1)}{4(d-3)} \ri]
\le(\f{\a}{\l_d}\ri)^{d-1}
\ri\}~t_{Pl}\,.
\la{decaytime1}
\ee
If the initial mass far exceeds the Planck mass, i.e.\ $m_i\gg1$, the last term
inside the curly brackets can be ignored. Using Eq.~(\ref{decayt}), we find
\be
\f{\D \t_1}{\t_0} \equiv \f{\t_{1}-\t_0}{\t_0}
= -\f{{\cal D}(d-1)\a^2}{4(d-3) \l_d^2}~m_i^{-2/(d-3)}\,.
\la{prop1}
\ee
The GUP-corrected decay time is smaller than the semiclassical decay time. The
GUP-corrected multiplicity is obtained from Eq.~(\ref{multi}) with
$S_0=S_{GUP}$. Table \ref{numbertable1} shows the GUP-corrected parameters for
two typical BHs produced at particle colliders or in UHECRs, with initial mass
equal to $5M_{Pl}$ and $10M_{Pl}$. The first row gives the standard Hawking
parameters. The GUP effects on the thermodynamic parameters increase as the
minimum BH mass becomes larger. It is interesting to note that decay time,
entropy, and multiplicity are drastically reduced when $\alpha$ approaches
unity. In the limiting case of a BH with initial mass $5M_{Pl}$ and GUP
parameter $\alpha=1$, the minimum BH mass coincides essentially with the
initial BH mass, and the BH does not evaporate. In contrast to the standard
theory, the GUP-corrected entropy shows that a BH with a mass five times the
fundamental Planck scale is not a classical object; quantum effects become
manifest at an earlier stage of the BH evaporation phase than was predicted by
the semiclassical Hawking analysis \cite{reviews}.  Therefore, GUP corrections
have important consequences on the BH phenomenology in particle colliders and
in UHECR airshowers. 

\begin{table*}[t]
\caption{GUP-corrected thermodynamic quantities for two ten-dimensional BHs
with mass $m=5$ and 10 (in fundamental units). The values in brackets
give the percentage deviation from standard Hawking quantities.}
\begin{center}
$m=5$
\vskip 6pt
\begin{tabular}{|c||c|c|c|c|c|c|}
\hline &~Minimum~mass~&~Initial~Temperature~&Final Temperature~&~Decay time~&~Entropy~&~Multiplicity~\\
\hline
$\alpha=0$ &  -- &  .553 &$\infty$&  .334 &  7.92 &  3 \\
$\alpha=0.5$ & .037 & .591 (+7\%)&2.23& .233 (-30\%)& 7.18 (-9\%)& 2
(-33\%)\\
$\alpha=1.0$ & 4.73 & .981 (+77\%)&1.11& .002 (-99\%)& .269 (-97\%)& 0
(-100\%)\\
\hline
\end{tabular}\vskip 24pt
$m=10$
\vskip 6pt
\begin{tabular}{|c||c|c|c|c|c|c|}
\hline &~Minimum~mass~&~Initial~Temperature~&Final Temperature~&~Decay time~&~Entropy~&~Multiplicity~\\
\hline
$\alpha=0$ &  -- &  .500 &$\infty$&  .814 &  17.5 &  6 \\
$\alpha=0.5$ & .037 & .529 (+6\%)& 2.23& .610 (-25\%)& 16.2 (-7\%)& 5
(-17\%)\\
$\alpha=1.0$ & 4.73 & .696 (+39\%)&1.11& .100 (-88\%)& 6.66 (-62\%)& 2
(-66\%)\\
\hline
\end{tabular}
\end{center}
\label{numbertable1}
\end{table*}

\section{Entropy Corrections Due to Thermodynamic Fluctuations} 
\la{entcorr}
Thermodynamic systems (including BHs) undergo small thermal fluctuations from
equilibrium which affect thermodynamic quantities. In this section we calculate
the corrections to entropy and Hawking temperature. We have seen that the GUP
corrections affect the Hawking temperature of the BH while the BH energy
remains constant. Therefore, we consider fluctuations around the equilibrium
temperature instead of the equilibrium energy. This leads to a decrease in the
black hole entropy. Note that the Bekenstein-Hawking entropy is identified with
the canonical entropy of the system\footnote{If fluctuations around the
equilibrium energy were considered, the Bekenstein-Hawking entropy could be
identified with the microcanonical entropy \cite{micro}.}. Let us consider a
canonical ensemble with partition function \cite{Das:2001ic,Das:2002ce}:
\be
Z(\b)
= \int_0^\infty \r(E) e^{-\b E} dE\,,
\label{part1}
\ee
where $\b=1/T$ is the inverse of the temperature. The density of states can be
obtained from Eq.~(\ref{part1}) by the inverse Laplace transform (at fixed
$E$) 
\be
\r(E) = \f{1}{2\p i} \int _{c-i\infty}^{c+i\infty} Z(\b) e^{\b E} d\b
= \f{1}{2\p i} \int_{c-i\infty}^{c+i\infty} e^{S(\b)}d\b\,,
\la{density1}
\ee
where
\be
S(\b) = \ln Z(\b) + \b E\,.
\label{basic}
\ee
To evaluate the complex integral in Eq.~(\ref{density1}) by the method of 
steepest descent, we expand $S(\b)$ around the saddle point $\b_0 (=1/T_0)$,
where $T_0$ is the equilibrium temperature. Also using the fact that $S_0' =
(\pa S(\b)/\pa\b)_{\b=\b_0}$, we get \cite{Das:2001ic,Bhaduri:2003kv}:
\be
S = S_0 + \f{1}{2}~(\b - \b_0) ^2 S_0'' + \cdots\,,
\label{ent1}
\ee
where $S_0=S(\beta_0)$ and $S''_0=(\pa^2 S(\b)/\pa \b^2)_{\b=\b_0}$.
Substituting Eq.~(\ref{ent1}) in Eq.~(\ref{density1}), the density of states is
\ba
\r(E) &=& \f{e^{S_0}}{2\p i} \int_{c-i\infty}^{c+i\infty}
e^{(\b - \b_0)^2 S_0''/2 } d\b \la{ent23} \nn\\
&=& \f{e^{S_0}}{\sqrt{2\p S''_0}}\,.
\la{corr0}
\ea
The corrected entropy is given by the logarithm of the density of states
$\r(E)$:
\be
S = \ln \r(E) = S_0 - \f{1}{2} \ln S_0'' + ~\mbox{(higher order terms)}\,.
\la{corr1}
\ee
From $E = - (\pa\ln Z(\b)/\pa\b)_{\b_0}$ and the definition of specific heat,
$C = (\pa E/\pa T)_{\b_0}$, $S''(\b)$ can be written as
\ba
S''(\b) &=& {1\over Z}({\pa^2 Z(\b)\over {\pa \b^2}})-
{1\over Z^2}({\pa Z\over \pa \b})^2  \nn \\
&=& \langle E^2\rangle-\langle E\rangle^2 \nn \\
&=& C~T^2\,.
\label{esci}
\ea
Equation (\ref{esci}) shows that a non-vanishing $S_0''$ is a consequence of
thermal fluctuations.  Substituting Eq.~(\ref{esci}) in Eq.~(\ref{corr1}), we
obtain
\be
S = \ln \r = S_0 - \f{1}{2} \ln \le( C~T^2 \ri) + \cdots\,.
\la{corr3}
\ee
The above formula applies to any thermodynamic system in equilibrium.  In
particular, when applied to BHs, $T$ is the Hawking temperature. For example,
for a non-rotating three-dimensional BTZ BH, both $T_H$ and $C$ are proportional
to the BH entropy $S_0$. In this case, Eq.~(\ref{corr3}) reads
\ba
S = \ln \r &=& S_0 - \f{3}{2} \ln S_0 + \cdots\,.
\la{btz4}
\ea
Similarly, for an AdS-Schwarzschild BH in $d$-dimensions, it can be shown that
(see Ref.~\cite{Das:2001ic})
\be
S = S_0 - \f{d}{2(d-2)} \ln S_0  + \cdots\,.
\la{scadscorr2}
\ee
Although Eq.~(\ref{corr3}) is not applicable directly to the Schwarzschild BHs
because of its negative specific heat, the entropy corrections can be shown to
be logarithmic by either assuming a small cosmological constant,
or by putting the BH into a finite box. The result is:
\be
S_{Thermo} = S_0 - k \ln S_0~\,,
\la{corr2}
\ee
where $k$ is a positive constant of order unity.  We will apply these results
to brane world BHs in the next section.
\section{Corrections to BH decay rate due to thermodynamic
fluctuations}
\la{fluct}
The corrected Hawking temperature is obtained from the first law of BH
thermodynamics. The result is:
\be
T_H'' = \f{(d-3)}{4\p \l_d} m^{-1/(d-3)}~\le[1+
\f{k(d-2)}{4\p\l_d}~m^{-(d-2)/(d-3)}+\dots\ri]~M_{Pl}c^2\,.
\la{taylor2}
\ee
Equation (\ref{taylor2}) gives the first-order correction to the Hawking
temperature
\be
\D T_{Thermo} \equiv T_H'' - T_H = \f{k(d-2)(d-3)}{16\p^2\l_d^2}~
m^{-(d-1)/(d-3)}~M_{Pl}c^2\,.
\la{delta2}
\ee
The correction to the BH entropy follows from Eq.~(\ref{corr2}):
\be
\D S_{Thermo}  = - k \ln S_0\,. 
\ee
The multiplicity is also reduced, and is now given by Eq.~(\ref{multi}) with
$S_0 \rightarrow S_{Thermo}$. Similarly to the GUP corrections, thermodynamic
fluctuations reduce the number of degrees of freedom of the BH. 
Taking the ratio of Eq.~(\ref{delta1}) and Eq.~(\ref{delta2}), we find 
\be
\f{\D T_{GUP}}{\D T_{thermo}} = \le[\f{\p\a^2}{\l_d k(d-2)} \ri]
~ m^{(d-4)/(d-3)}~\,.
\la{ratio}
\ee
The GUP and the thermal fluctuation corrections are of the same order for
$d=4$. In $d > 4$, the situation is more complicated. If $m\gg 1$, the GUP
corrections far exceed the corrections due to thermodynamic fluctuations.
However, when $m \approx 1$, i.e.\ near the end stage of evaporation, the rates
are comparable. The first-order corrected specific heat is:
\be
{\cal C} ={\cal
C}_0\left[1-\frac{k(d-1)(d-2)}{4\p\l_d}m^{-(d-2)/(d-3)}\right]\,,
\la{sh2}
\ee
where ${\cal C}_0$ is given by Eq.(\ref{c0}). The first-order specific heat
vanishes for the non-zero value of $m$
\be 
m_0 = \le[\f{k(d-1)(d-2)}{4\p\l_d}\ri]^{(d-3)/(d-2)}\,.
\ee
This suggests that the BH  becomes thermodynamically stable when the BH mass
reaches $m_0$. However, this conclusion should be interpreted with care; the
thermodynamic fluctuations of a BH with mass $m\sim m_0$ are large and the
first-order approximation (\ref{corr2}) breaks down. The Stefan-Boltzmann law
is obtained from Eq.~(\ref{taylor2}):
\be
\f{dm}{dt} =
- \f{\m}{t_{Pl} m^{2/(d-3)}}
\le[ 1 + \f{k{\cal D}(d-2)}{4\p\l_d}m^{-(d-2)/(d-3)}+\dots\ri]\,.
\la{rate6}
\ee
The first-order correction to the Stefan-Boltzmann law due to thermal
fluctuations is positive:
\be
\D\le(\f{dm}{dt}\ri)/\le(\f{dm}{dt}\ri)_0=
\f{k{\cal D}(d-2)}{4\p\l_d}~m^{-(d-2)/(d-3)}\,.
\ee
Integrating Eq.~(\ref{rate6}) we obtain the expression for the time decay
\be
\t_2 = \m^{-1} \le( \f{d-3}{d-1} \ri) m_i^{(d-1)/(d-3)}
\le[1 -
\f{k {\cal D}(d-1)(d-2)}{4\p\l_d}~m_i^{-(d-2)/(d-3)}
+\dots \ri]t_{Pl}\,.
\la{decaytime2}
\ee
Similarly to the GUP case, it can be easily verified that the BH takes less
time to decay when fluctuation corrections are taken into account: 
\be
\f{\D \t_2}{\t_0} \equiv \f{\t_{2}-\t_0}{\t_0}
= -\f{k{\cal D}(d-1)(d-2)}{4\p\l_d}~m_i^{-(d-2)/(d-3)}\,.
\la{prop2}
\ee
The decay time is dramatically reduced by thermal fluctuations. Equation
(\ref{decaytime2}) implies a relation between the thermal fluctuation threshold
$m_0$ and the BH initial mass. By imposing $\tau_2>0$ we have
\be
m_i>{\cal D}^{(d-3)/(d-2)}m_0\equiv m_0'\,.
\la{thmass}
\ee
BH with initial mass smaller than $m_0'$ form in a regime where thermal
fluctuations dominate. A careful study of their thermodynamic properties should
include higher-order terms in the expansion (\ref{ent1}).  Since the
thermodynamic fluctuations prevent the analytical evaluation of the integral in
Eq.~(\ref{ent23}), numerical techniques may have to be used to get accurate
estimates of the thermodynamic quantities. In any case, our analysis shows that
semiclassical Hawking theory is inadequate for the the description of these
black holes.  Note that for $k=0.5$ ($1$), $m_0'=10.25$ ($18.8$). Thermal
fluctuations cannot be neglected in particle collider or UHECR BH events.
\section{Discussion}
\la{concl}
We have examined the effects of the GUP and small thermal fluctuations on
temperature, decay rate, and entropy of microscopic BHs. Although these effects
are small under most circumstances, they can be significant in BH production at
the TeV scale, where the BH mass is expected to be of the order of the
fundamental Planck mass. The GUP and the thermal fluctuation corrections
increase the BH temperature, and decrease decay time, entropy, and multiplicity
of the evaporation phase: Quantum BHs are hotter, shorter-lived, and tend to
evaporate less than classical BHs. The results described here are applicable to
the ADD as well as the RS brane world scenarios \cite{add,rs}. 

Under the most favorable circumstances, the semiclassical cross section for BH
formation at the TeV scale reaches hundreds of pb for proton-proton collision
at the LHC, and millions of pb for neutrino-nucleon collision in the
atmosphere. According to the semiclassical scenario, the Hawking evaporation
mechanism will allow detection of microscopic BHs with mass of the order of a
few Planck masses in next-generation particle colliders and UHECR detectors.
However, our results seem to suggest that the semiclassical description could
be inaccurate for this kind of events.

Firstly, a shorter lifetime implies that the quantum BH may not behave like a
well-defined resonance. Secondly, the classical picture breaks down if the
degrees of freedom of the BH, i.e.\ its entropy, is small. The semiclassical
entropy has been widely used in the literature to measure the validity of the
semiclassical approximation. When the entropy is sufficiently large, the BH can
be considered as a classical object \cite{reviews}. If this is the case, i) the
BH evaporation phase is described by a thermal spectrum with Hawking
temperature; ii) the BH cross section for elementary particles is well
approximated by the geometrical cross section; and iii) the total cross section
for composite particles (e.g.\ nucleons) is obtained by integrating the
geometrical cross section over the structure functions. A BH with mass equal to
few Planck masses is usually assumed to have entropy above the threshold of
validity of the classical description. However, the reduction in entropy by GUP
and thermal fluctuation effects increases this threshold. Therefore, it may not
be appropriate to treat these BHs as classical objects. (See also
Ref.~\cite{criticisms}). Thirdly, GUP physics implies the existence of a
minimum BH mass given by Eq.~(\ref{minmass}). The existence of a minimum mass
increases the lower cutoff for BH formation, thus reducing the rate of BH
events. Even if a detectable signal is produced during the BH decay phase, it
could prove very difficult to distinguish it from the background. Finally, GUP
and thermodynamic fluctuations further decrease the already-weak lower bounds
on the fundamental Planck scale that follow from the nonobservation of BH
events up to date \cite{limits}.

Let us conclude with a list of open problems and possible future research
topics. It would be interesting to compute the effects of small residual charge
$Q$ and angular momentum $J$ on the above results. In presence of nonzero
charge and angular momentum, the corrections to the thermodynamic quantities
are expected to depend on $J$ and $Q$. However, the precise form of the
corrections is yet to be determined. It would also be interesting to examine
the relation of the GUP and thermodynamic corrections to other corrections that
have been predicted for TeV-scale BHs (see e.g. Refs.~\cite{reviews},
\cite{criticisms}). A detailed single-event analysis of particle collider and
UHECR events when GUP and thermodynamic corrections are present, is also
missing. We plan to report on these and other issues elsewhere.

\vs{.4cm}
\no
\ul{{\bf Acknowledgements:}}

\vs{.2cm}
\no

We thank R.\ Maartens and M.\ Maggiore for interesting discussions. S.D.\ would
like to thank R. K.\ Bhaduri, A.\ Dasgupta, J.\ Gegenberg, V.\ Husain, P.\
Majumdar  and M.\ Walton for interesting discussions and comments. This work
was supported in part by the Natural Sciences and Engineering Research Council
of Canada and funds of the University of Lethbridge.

\end{document}